  \documentstyle[12pt]{article}
\catcode`@=11
\def\secteqno{\@addtoreset{equation}{section}%
\def\theequation{\thesection.\arabic{equation}}}
\catcode`@=12
\secteqno
\newcommand{\be}{\begin{equation}}
\newcommand{\ee}{\end{equation}}
\newcommand{\bea}{\begin{eqnarray}}
\newcommand{\eea}{\end{eqnarray}}
\newcommand{\bref}[1]{(\ref{#1})}
\newcommand{\slp}{/ {\hskip-0.27cm{p}}}
\newcommand{\slP}{/ {\hskip-0.27cm{P}}}
\newcommand{\slX}{/ {\hskip-0.27cm{X}}}
\newcommand{\sltp}{/ {\hskip-0.27cm{\tilde{p}}}}
\newcommand{\slPi}{/ {\hskip-0.27cm{\Pi}}}

\def\CM{{\cal M}}

\begin{document}
          {\hfill August~1,~ 1997}\par
	  {\hfill Toho-FP-9756}\par
	  {\hfill hep-th/9708001}
%\title{Covariant Quantization of Super-D-string}

\vskip 20mm
\begin{center}
{\bf \Large Covariant Quantization of The Super-D-string}
\vskip 10mm
%\author
{\large Machiko\ Hatsuda and Kiyoshi\ Kamimura$^\dagger$}\par
%}\address{
\medskip
{\it Department of Radiological Sciences,
Ibaraki Prefectural University of Health Sciences,\\
\ Ami\ Inashiki-gun\ Ibaraki\ 300-03, Japan \\
%}\address{
$^\dagger$ Department of Physics, Toho University,
\ Funabashi\ 274, Japan
}
\medskip
\date{\today}
\end{center}
%\maketitle
\vskip 10mm
\begin{abstract}
We present the covariant BRST quantization of the super-D-string. 
The non-vanishing supersymmetric 
U(1) field strength ${\cal F}$ is essential
for the covariant quantization of the super-D-string
as well as for its static picture.  
A SO(2) parameter parametrizes a family of local supersymmetric
 (kappa symmetric) systems including
the super-D-string with ${\cal F}\ne 0$ and 
the Green-Schwarz superstring with ${\cal F}= 0$.
We suggest that $E^1$ (canonical conjugate of U(1) gauge field)
plays a role of the order 
parameter in the Green-Schwarz formalism: 
the super-D-string exists for $E^1 \ne 0$ 
while the fundamental Green-Schwarz 
superstring exists only for $E^1=0$.

\end{abstract}
%\pacs{}
\par

\newpage
\setcounter{page}{1}

\parskip=7pt
%%%%%%%%%%%%%%%%%%%%%%%%%%%%%%%%%%%%%%%%%%%%%%%%%%%%
\section{ Introduction}
\indent

D-branes play important roles to study non-perturbative aspects of 
superstring theories \cite{Pol}. Especially the SL(2,Z) duality found 
in the low energy effective theory of the
 type IIB superstring theory \cite{Shsl} 
relates the fundamental superstring and a series of soliton solutions as 
the SL(2,Z) multiplet. The fact that super-D-brane states belong to 
the same multiplet as those of type II fundamental superstring tells 
they share the same symmetry structures.
Intensive studies of dualities are establishing
physical pictures of D-branes \cite{Pol,Shsl,Witt,Shdd}.
The D-string ( D1 brane )
is required as the S dual object of the type IIB superstring and 
has properties: 
(1) it has the R-R charge, 
(2) it is the BPS saturated state,  
(3) its tension is scaled as a representation of SL(2,Z),
(4) it is consistent with the static gauge. 
The action for super-D-branes was written down
in the manifest symmetric form
under  global space-time supersymmetry and 
$\tau$, $\sigma$ reparametrization and the local supersymmetry
(kappa symmetry) 
\cite{Cw,Shgf,Shkp,BT}.
Since the guiding principle of construction of actions is
the local symmetries, the super-D-string and the Green-Schwarz 
superstring have similar symmetry structures though 
their physical pictures are quite different.
In this paper, we start with the super-D-brane action \cite{Cw,Shgf,Shkp,BT} 
and examine how above physical pictures of D-branes
are arisen.

There is also an interesting issue of  
covariant quantization of super-D-branes as
space-time supersymmetric objects.
Despite of simple description of the Green-Schwarz (GS) superstring \cite{GS}
the covariant quantization without infinite number of fields
has been a long standing problem. 
The origin of the difficulty of the covariant quantization of GS superstring
is existence of massless point-like ground states.
It is suggested that such massless ground states do not appear 
once we accept the super-D-brane picture \cite{RK}. 
This massiveness 
condition is satisfied if the static gauge were taken \cite{Shgf}.
The real question is whether 
the static gauge can be chosen and the massiveness condition is
 satisfied for the D-brane states especially 
for the ground states.  
To answer it
we rather begin by the action and clarify the condition
for the static and massive picture of super-D-branes. 

The physical picture of D-branes is the one of ``fat strings"
described by the partial spontaneous breaking of the
global supersymmetry algebra
\cite{HP}.
A lying string breaks some of translational symmetries and supersymmetries,
and it gives rise to goldstone coordinates and goldstinos
respectively.
In order to make Lorentz covariant action,
extra coordinates 
associated with unbroken symmetry generators
are introduced in the gauge invariant way,
so that extra coordinates are set to be 
redundant.
For the super-D-string system,
covariance of the gauge, $\theta_1=0 $, 
reflects covariance of the Nambu-Goldstone fermion,
$\theta_2$.
Due to the Wess-Zumino action \cite{Cw,Shgf,Shkp,BT},
the global supersymmetry 
algebra allows the central extension \cite{Pol}
which leads to the BPS condition
representing the spontaneous symmetry breaking.
Since the global supersymmetry algebra is similar to
 the local supersymmetry algebra,
the central extension enables us to separate
fermionic constraints in covariant and irreducible way.
In this paper we confirm this beautiful consistency in the canonical language.
We also reexamine the covariant gauge from the viewpoint of 
 the equation of motion and the boundary condition.

The plan of this paper is the following: In section 2, we perform canonical 
analysis and the BRST quantization of the super-D-particle. Although it has 
been examined in detail in \cite{RK} we review it for the later application. 
Especially the structure of the irreducible covariant separation of the 
first class and the second class constraints is clarified. 
The
BPS bound is derived from the N=2 global supersymmetry algebra. 
One half of supersymmetry is realized as usual
 by using the Nambu-Goldstone fermion, and another half is realized
 rather trivially. 
This is explained in terms of the 
Dirac stared variables.
In section 3, the super-D-string action is analyzed in the canonical formalism.
The singularity which would prevent the covariant quantization
is carefully examined.
In section 4, the BRST charge of the super-D-string is constructed
and the covariant gauge fixed actions are obtained. 
We clarified how the super-D-string overcomes difficulties of the covariant quantization of the superstring, such as how the ghost for ghost does not appear. 
The global supersymmetry algebra is calculated
 and the BPS property is explained.
In section 5, we present a family of local supersymmetric 
actions parametrized by a SO(2) parameter and 
the relation between the Green-Schwarz superstring and 
the super-D-string is clarified. 
It is also noted that the general background makes the SUSY central 
charges SL(2,R) covariant representation.

%%%%%%%%%%%%%%%%%%%%%%%%%%%%%%%%%%%%%%%%%%%%%%%%%%%%
\section{Covariant quantization of the super-D-particle}
\indent

The action of the super-D-particle is
given by \cite{Shgf} 
\bea
S&=&-T\int d\tau \sqrt{-{\rm det}(G+{\cal F})}+
T\int \Omega_{(1)}\nonumber\\ &=&\int 
d\tau ~T~[-\sqrt{-\Pi_0^2}~+~\bar{\theta}\Gamma_{11}\dot{\theta}~] 
\eea
where $\Pi_0^m=\dot{X}^m-\bar{\theta}\Gamma^m\dot{\theta}$
is the super symmetric velocity.
The Majorana-Weyl spinors $\theta_A (A=1,2) $ have opposite 
chiralities; $\Gamma_{11}~\theta_A~=~(-1)^{A-1}~\theta_A~$.
Although the system is examined in some detail by Kallosh \cite{RK}
we give the canonical analysis here with some remarks.

In the canonical formalism it follows the primary constraints
\bea
h&\equiv& \frac{1}{2T}( p^2+T^2)~ =~0\label{Lpar}\\
 f&\equiv& \zeta~+~\bar{\theta}~(~\slp~-~T~ \Gamma_{11})~=~0\label{fpar}
\eea
where 
$\zeta_A$, with $\zeta_A\Gamma_{11}=(-1)^{A-1}\zeta_A$, 
are canonical momenta conjugate to $\theta_A$.
The first one is the mass shell condition that the D-particle
has its mass $T$. The second is fermionic constraints
satisfying following algebra
\bea
\{{f}_{A,\alpha},{f}_{B,\beta}\}=
2(C\Sigma)_{A,\alpha\ B,\beta}
\label{fercon}
\eea
with
\bea
\Sigma_{AB}&=&
(\slp-T\Gamma_{11})
=\left(\begin{array}{cc}
-T&\slp\\
\slp&T
\end{array}\right)
\quad.\label{Sigma}
\eea
Here the chiral representation is chosen such that
 $\Gamma_{11}$ is diagonal and the charge conjugation matrix, $C$,
and $\Gamma_{\mu}$ are off-diagonal.
As in the case of massless superparticle, 
it is nilpotent on the constraint \bref{Lpar} 
as a matrix 
\bea
\sum_{B=1,2}
\Sigma_{AB}\Sigma_{BC}\approx 0\label{fer1}\quad.
\eea
However each block of the matrix $\Sigma$ is not nilpotent 
so it has non-zero determinant
in contrast with the superparticle case,
\bea  
\det~(\Sigma_{AB})~\ne 0 \label{fer2}\quad .
\eea 
Irreducible and covariant separation of fermionic constraints
is possible by using (\ref{fer1}) and (\ref{fer2}).
One half of eigenvalues
is zero while another half is non-zero.
In the rest frame $\Sigma\rightarrow T\Gamma_0(1+\Gamma_0\Gamma_{11})$
where $(1+\Gamma_0\Gamma_{11})/2$ is a projection operator.
Hence the rank of $\Sigma$ is one half of its maximum, 32/2. 
Therefore one half of the fermionic constraints, $f_{A,\alpha}$, is 
the first 
class and another half is the second class. 
The fermionic first class constraints are
\be
\tilde{f}_2\equiv f\Sigma\frac{1-\Gamma_{11}}{2}
=f(-T\Gamma_{11})(1-\gamma^{(0)})\frac{1-\Gamma_{11}}{2}
=\zeta_1\slp+T\zeta_2+\bar{\theta}_1(p^2+T^2)=0\quad ,\label{pf1}
\ee
where the projection factor $(1-\gamma^{(0)})$
is the canonical version of the projection operator of Schwarz \cite{Shgf}.
The second class constraints are 
\be
f_1\equiv f\frac{1+\Gamma_{11}}{2} =
\zeta_1+\bar{\theta}_1\slp-T\bar{\theta}_2=0
\label{pf2}\quad.
\ee
The algebra of first class constraints and second class constraints is
calculated as;
\bea
\{\tilde{f}_2,\tilde{f}_2\}
=-4TC\slp h 
~~,~~
\{f_1,f_1\}=2C\slp~~,~~
\{\tilde f_2,f_1\}=-2TC h\quad .
\eea
The Dirac bracket for the second class constraints, $f_1=0$, 
is defined by 
\bea
\{ A,B\}_D
&=&\{ A,B\} -\{ A,f_1\}\frac{\slp C^{-1}}{2p^2}
\{ f_1,B\} \quad .
\eea

\medskip

The covariant separation of fermionic constraints into
(\ref{pf1}) and  (\ref{pf2}) enables us to construct
 the covariant path integral. 
The second class constraints are built in the path integral measure \cite{FF}: 
\bea
&&Z=\int d\mu_{\rm can}
 \ \delta(f_1)(\det\{f_1,f_1\})^{-1} \ \exp i\left[\int  
\left(p\dot{X} +\zeta_A\dot{\theta}_A+b\dot{c}+\beta_2\dot{\gamma}_2
- H_{gf+gh} \right)\right] \nonumber\\ 
&&~~~~~d\mu_{\rm can}={\cal D}[\zeta_A , \theta_A,  p, X;
c,b,\gamma_2,\beta_2]\ 
\quad .\label{gfp}
\eea
Where the gauge fixed and ghost part of Hamiltonian is
\be
H_{gf+gh}=\{ \Psi ,Q_B\}_D
\label{Hgfgh}\ee
with the gauge fixing function, $\Psi$, and the BRST charge, $Q_B$. 

The BRST charge is obtained 
as
\be
Q_{B,{\rm min}}=hc+\tilde{f}_2\gamma_2+2bT
\bar{\gamma}_2\slp\gamma_2
\ee
with nilpotent property; 
$
\{ Q_{B,{\rm min}},Q_{B,{\rm min}}\}_D=0$.
Since we have irreducible constraints infinite number of ghosts 
are not required unlike the superparticle case. 
In order to perform gauge fixing, a bosonic auxiliary field
 and a fermionic auxiliary canonical pair $\psi$ and $\pi_{\psi}$
are introduced.
The total BRST charge is given by
\be
Q_B=Q_{B,{\rm min}}+\pi_g\tilde{c}+\pi_{\psi}\tilde{\gamma}_2
\ee
with nilpotent property;
$
\{ Q_{B},Q_{B}\}_D=0\quad$.

Corresponding to the covariant gauge, $\theta_2=0$, 
the gauge fixing function is taken as,
\be
\Psi= bg-\beta_2 \psi_2
-\tilde{\beta}_2\theta_2\label{fix}\quad .
\ee
Here the reparametrization gauge is kept unfixed 
in order to focus on fermionic gauge fixing procedure.
Auxiliary field $\chi$, $\eta_1$ and $\rho_1$ are introduced to exponentiate 
the contribution of second class constraints in the measure. 
Then the path integral becomes
\bea
Z_{\rm }=\int d\mu_{\rm can}{\cal D}[ \chi_1, \eta_1, \rho_1; g,\psi,\pi_{\psi}, \tilde{\gamma}_2,\tilde{\beta}_2] \ \exp i\int {\cal L} \
\eea
\bea 
{\cal L}&=&f_1\chi_1 +\bar{\eta}_1\frac{\slp}{2T^2}\rho_1+
p(\dot{X}-\bar{\theta}\Gamma\dot{\theta})-gh
+\tilde{f}_2\psi_2 +\pi_{\psi}\theta_2\nonumber\\&&
+b\dot{c}+b\frac{4}{T}\bar{\gamma}_2\slp
\psi 
+\beta_2\dot{\gamma}_2
+\beta_2\tilde{\gamma}_2+\tilde{\beta}_2\gamma_2\quad .
\eea
Most variables can be integrated out
and it ends up with just overall constant. 
There remain integrations over physical variables, $X$ and $\theta_1$, and 
reparametrization ghosts, $b$ and $c$:
\bea
Z_{\rm }
& = & \int {\cal D} [X,p, \theta_1, g,{b},c]
\ \exp \left[i\int \left(p(\dot{X}-\bar{\theta}_1\Gamma\dot{\theta}_1)
-gh +{b}\dot{c}\right)\right] \label{gfcan}\\
& = & \int {\cal D} [X, \theta_1, g,{b},c]
\ \exp \left[i\int \left(-\frac{T}{2g}(\dot{X}-\bar{\theta}_1\Gamma\dot{\theta}_1)^2
-\frac{g}{2}T +{b}\dot{c}\right)\right] \label{A2}\\
&=&\int {\cal D} [X, \theta_1, {b},c]\ \exp \left[i\int \left(-T\sqrt{-(\dot{X}-\bar{\theta}_1\Gamma\dot{\theta}_1)^2}
 +{b}\dot{c}\right)\right] \label{A1} 
\eea
The canonical action becomes second order form action (\ref{A2}) by 
integrating out the canonical momentum $p$, and the second order form 
action reduces to the square root
type action by integrating out the einbein $g$.
Although there exists dynamical ghost parts, they never appear
as long as physical states matrix elements are taken.

The gauge fixed action in the conformal gauge is obtained by setting 
$g=1$ in the gauge fixing function in (\ref{fix}), 
\bea
I=\int \left(p(\dot{X}-\bar{\theta}_1\Gamma\dot{\theta}_1)
-\frac{1}{2T}(p^2+T^2) \right)\label{gfconp}\quad.
\eea
Then equations of motion in this gauge
are calculated as 
\bea
T\ddot{X}=0 \label{eqb}\quad ,\quad
\dot{\theta_1}=0\quad \label{eqf} .
\eea
Although $X$ and $\theta_1$ satisfy free field equations,
the action can not be written in simple free form 
for the manifest supersymmetry and the Lorentz covariance.
If the fermionic part of action is written in 
free form \cite{RK}, coordinates transform non-covariantly 
under both the Lorentz and the global supersymmetry transformations. 

\medskip
\medskip
This system has the N=2 global supersymmetry whose charges are given by
\bea
Q_A\epsilon_A=\zeta\epsilon-\bar{\theta}(\slp-T\Gamma_{11})\epsilon
\quad.
\eea
They satisfy the SUSY algebra with the central extension given by 
\be
\{Q_{A,\alpha},Q_{B,\beta}\}_D=-2(C\Sigma)_{A,\alpha\ B,\beta}\quad.
\label{227}
\ee
The use of the Dirac bracket does not change the algebra
 calculated with the Poisson bracket since Poisson brackets between
global SUSY charges and fermionic constraints are zero, $\{Q_A,f_B\}=0$.
Since the right hand side of \bref{227}
 is the same as $\Sigma$ in \bref{Sigma},
one half of eigenvalues is zero. 
If we choose the linear combination of these supercharges as
\bea
\left\{\begin{array}{ccl}
\tilde{Q}_1&=&Q_1\\
&&\\
\tilde{Q}_2&=&Q_2+Q_1{\slp}/{T} 
\end{array}\right.\quad,
\eea
their Dirac brackets become
\bea
\{\tilde{Q}_A,\tilde{Q}_B\}_D
=-2
\left(\begin{array}{cc}
C\slp&2Ch\\&\\
2Ch&-\frac{2}{T}C\slp h
\end{array}\right)
\quad .\label{D0BPS1}
\eea
Here $h$ is the constraint in \bref{Lpar}.
In the canonical quantization procedure the Dirac bracket is
replaced by the commutator;
$\{ q,p\}_D=1\rightarrow [q,p]=i$. 
Using the above SUSY algebra, the BPS bound is obtained: 
\be
\parallel \tilde{Q}_2^{\alpha}|\ast \rangle \parallel^2=\frac{1}{2}\langle \ast|[\tilde{Q}_2^{\alpha},\tilde{Q}_2^{\alpha}]_+|\ast\rangle
=\frac{4}{T}p_0(p_0-T)(p_0+T)\parallel |\ast \rangle \parallel^2\geq 0\quad ,
\ee
in the rest frame ($p_0>0$) with $C=i\Gamma_0$.
The $\tau$ reparametrization constraint, $h=0$ in (\ref{Lpar}),
leads to that the super-D-particle is the BPS saturated state.
For the BPS saturated state, $p_0=T$, the SUSY algebra becomes
\bea
[\tilde{Q}_A,\tilde{Q}_B]_+
 = 
\left(\begin{array}{cc}
2T&0\\
0&0
\end{array}\right)\label{BPSD0}\quad.
\eea
This shows that 
 $\tilde{Q}_1^{\alpha}$ invariance is spontaneously broken
and $\theta_1$ is recognized as the Nambu-Goldstone fermion.
On the other hand,
 $\tilde{Q}_2^{\alpha}$ invariance is preserved. 
In fact  $\tilde{Q}_2$ is equivalent to the gauge generator
\be
\tilde{Q}_2=\frac{1}{T}\tilde{f}_2-4\bar{\theta}_1h\quad.\label{tq20}
\ee
Therefore $\tilde{Q}_2$ generates rather the local supersymmetry than the global supersymmetry.
The super-D-particle system represents one half of the N=2 global 
supersymmetry generated by $\tilde{Q}_1$
\be
\{\tilde{Q}_1,\tilde{Q}_1\}=-2C\slp\quad.\label{Q1Q1}
\ee

Indeed the gauge fixed actions, (\ref{gfcan}),
(\ref{A2}) and (\ref{A1}), have manifest global supersymmetry 
generated by $Q_1=\tilde{Q}_1$ as \bref{Q1Q1}, 
but they do not represent
 another half of SUSY generated by $Q_2$. 
The $Q_2$ invariance seems to contradict with the gauge fixing of 
the local supersymmetry, $\theta_2=0$: 
$
\{\theta_2, Q_2\epsilon_2\}=\epsilon_2\ne 0 
$.
However it is just superficial and the use of the Dirac bracket with the 
correct second class constraints set
resolves this.
After gauge fixing the Dirac bracket is modified by taking into 
account the set of second class constraints,
$f_1=f_2=\theta_2= 0$,
\bea
\{A,B\}_{D2}&=&\{A,B\}_D
-\{A,\theta_2\}\frac{TC\slp C^{-1}}{p^2}\{f_1,B\}
-\{A,\theta_2\}\{f_2,B\}\nonumber\\&&
+\{A,f_1\}\frac{T\slp}{p^2}\{\theta_2,B\}
-\{A,f_2\}\{\theta_2,B\}
\label{D21}\quad.
\eea
Hence the $Q_2$ transformation keeps the gauge condition 
$\theta_2=0$ invariant,  
$
\{\theta_2, Q_2\epsilon_2\}_{D2}=\epsilon_2-\epsilon_2= 0$.
In other words, the $\theta_2=0$ gauge is conserved
with using the local supersymmetry as discussed in \cite{Shgf}.
The $Q_2$ invariance of the gauge fixed action (\ref{gfconp}) 
can be checked explicitly by using with 
 non-linear transformation rules 
\bea
\delta_2\theta_1=\{\theta_1,Q_2\epsilon_2 \}_{D2}=
\frac{T\slp}{p^2}\epsilon_2
\quad ,\quad
\delta_2 X=\{X,Q_2\epsilon_2 \}_{D2}=T\bar{\theta}_1\Gamma\frac{\slp}{p^2}
\epsilon_2
,\nonumber
\eea
\bea
\rightarrow \delta_2 I=\int 
[p\left( \delta_2 \dot{X}-\delta_2(\bar{\theta}_1\Gamma\dot{\theta}_1)
\right)]
=2\int \partial_0 (T\bar{\theta}_1\epsilon_2)=0 \quad \label{hfSUSY}\quad.
\eea
The situation becomes clearer
by introducing the Dirac * variable
\bea
A^{\ast} \equiv A-\phi_a ( \{ \phi,\phi \} )_{ab}^{-1} \{ \phi_b,A \}
\quad .
\eea
The $Q_1$ symmetry is realized as usual while 
the $Q_2$ symmetry is realized 
associating with the local supersymmetry:
\be
Q_1^{\ast}=Q_1 \quad ,\quad Q_2^{\ast}=Q_2-\frac{1}{T}\tilde{f}_2 \quad .
\ee

The true degree of freedom of the 
global supersymmetry is one half of it 
by comparing the transformation rule $\delta_2$ in \bref{hfSUSY}
with $\delta_1$;
\be
T\frac{\slp}{p^2}\epsilon_2=\epsilon_1
\quad .
\ee
This relation is nothing but the relation \bref{tq20},
$\tilde{Q}_2=Q_2+Q_1\slp/T\approx 0$.
Therefore the BPS condition, 
\bref{D0BPS1} and \bref{BPSD0}, is explained as follows:
In the $\theta_2=0$ gauge, one half of the global SUSY generated by $Q_2$
is a linear combination of $Q_1$
by using with the local SUSY generated by $\tilde{f}_2$.
Surviving spinor, $\theta_1$, represents the one half of the global SUSY
generated by $Q_1$ manifestly.

%%%%%%%%%%%%%%%%%%%%%%%%%%%%%%%%%%%%%%%%%%%%%%%%%%%%%%%%%%%%%%%%%%%%%%%%%%%%%
\section{ Canonical analysis of the super-D-string}
\indent

The action of the super-D-string \cite{Shgf} is given by
\bea
 S&=&-T\int d^2\sigma \sqrt{-{\rm det}(G+{\cal F})}+
T\int \Omega_{(2)}\nonumber\\ 
 &=&T\int d^2\sigma [-\sqrt{-G_F}-\epsilon^{\mu\nu}
\Omega_{\mu\nu}(\tau_1)] \quad
\label{sdact}
\eea
where
\bea
\begin{array}{rclcrcl}
G_F&=&{\det}(G_{\mu\nu}+{\cal F}_{\mu\nu}) 
   =G+{\cal F}_{01}^2 &,&
G&=&{\det}G_{\mu\nu} =G_{00}G_{11}-G_{01}^2  \\
G_{\mu\nu}&=&\Pi_{\mu}^m\Pi_{\nu,m}&,&
\Pi_{\mu}^m&=&\partial_{\mu}X^m-\bar{\theta}\Gamma^m\partial_{\mu}
\theta
\\
{\cal F}_{01}&=&\epsilon^{\mu\nu}[\partial_{\mu}A_{\nu}-\Omega_{\mu\nu}(\tau_3)
]&&
&&\end{array}\nonumber
\eea
and
\bea
\Omega_{\mu\nu}(\tau_1)&=&-\bar{\theta}\Gamma\tau_1\partial_{\mu}{\theta}
\cdot (\Pi_{\nu}+\frac{1}{2}\bar{\theta}\Gamma\partial_{\nu}{\theta})\nonumber
\\
\Omega_{\mu\nu}(\tau_3)&=&-\bar{\theta}\Gamma\tau_3\partial_{\mu}{\theta}
\cdot 
(\Pi_{\nu}+\frac{1}{2}\bar{\theta}\Gamma\partial_{\nu}{\theta})
\quad .
\eea   
Notations of the Dirac matrices and the N=2 spinor indices are followed 
from the reference \cite{Shgf}.  
The definition of the canonical momenta leads to the following primary 
constraints : 
\bea
h&\equiv&\frac{1}{2T_E}(\tilde{p}^2+(T_E)^2G_{11})=0 \label{LH}\\
t&\equiv&\tilde{p}\cdot\Pi_1 =0 \label{LT}\\
F&\equiv&\zeta+\bar{\theta}
(\sltp-\slPi_1\tau_E)
-\frac{1}{2}(
\bar{\theta}\Gamma{\theta}'\cdot
\bar{\theta}\Gamma\tau_E+
\bar{\theta}\Gamma\tau_E{\theta}'\cdot
\bar{\theta}\Gamma )=0\label{F}\\
E^0&=&0. \label{Gaussz}
\eea
Here  $p_m$, $\zeta$ and $E^\mu$ are the canonical momenta conjugate to 
$X^m$, $\theta$ and $A_\mu$ respectively and
\be
\tilde{p}
\equiv p+\bar{\theta}\Gamma\tau_E{\theta}' \quad\label{3.7} .
\ee
The $\tau$ matrices,
which act on the $N=2$ spinor indices of $\theta_A$ with the same chirality,
appear only through the combination of $\tau_E$  
\be
\tau_E\equiv E^1\tau_3+T\tau_1\quad,~~~~~{\rm and}~~~~~~\tau_E^2~=~T_E^2
\quad.
\ee
Here $T_E$ defined by
\be
(T_E)^2\equiv T^2+(E^1)^2,
\label{TE}
\ee
is regarded as the tension of the D-string.

The algebra of fermionic constraints 
\bea
\{{F}_{A,\alpha},{F}_{B,\beta}\}=
2(C\Xi)_{A,\alpha\ B,\beta}
\label{fercons}
\eea
with
\bea
\Xi_{AB}~\equiv~\sltp\delta_{AB}-\slPi_1(\tau_E)_{AB}  
=
\sltp
\left(\begin{array}{cc}
1&0\\
0&1
\end{array}\right)
-\slPi_1
\left(\begin{array}{cc}
E^1&T\\
T&-E^1
\end{array}\right)
\quad.\label{Xi}
\eea
As same as the super-D-particle case, (\ref{fer1}) and (\ref{fer2}), 
$\Xi$ has nilpotency as a matrix 
\be
\Xi^2=2T_Eh-2\tau_Et\approx  0
\label{Xis}
\ee
but each block of $\Xi$ has non-zero determinant.
It follows that the rank of $\Xi$ is one half of 32 
as long as $T_E$ and $\Pi_1^2$ are non-zero.

The canonical Hamiltonian becomes a sum of the primary constraints
in a similar form as the GS superstring \cite{HK}, 
\bea
{\cal H}&=&\int\ d\sigma [p_m\dot X^m+\zeta_{A,\alpha}\dot\theta^{A,\alpha}+
E^\mu\dot{A}_\mu~-~{\cal L}~]
\nonumber\\
&=&\int \ d\sigma [g_0 {h}+g_1{t}+F\chi
+E^0\dot A_0-(\partial_1E^1)A_0]\nonumber\\
&=&\int \ d\sigma [g_0 \tilde{H}+g_1\tilde{T}+F\Xi\tilde\chi
+E^0\dot A_0-(\partial_1E^1)A_0]\quad.\label{canHam}
\eea 
In the second line, bosonic parameters are 
\be
g_0=\frac{\sqrt{-G_F}}{G_{11}}\frac{T_E}{T} \quad ,\quad
g_1=\frac{G_{01}}{G_{11}}\quad .
\ee
The consistency condition of the primary constraints \bref{Gaussz}
leads to 
the U(1) Gauss law constraint,
\be
\partial_1E^1~=~0\quad,
\label{Gauss}
\ee
as the secondary constraint. No further condition appears.
The fermionic parameter,
$\chi$, is determined by the consistency condition of $F$ in \bref{F}
to be conserved under the time development.
Using with (\ref{fercons}), (\ref{Xi}) and (\ref{Xis}) 
the consistency condition,
\be
\Xi~[~\chi~+~g_0~\frac{\tau_E}{T_E}~\theta'~-~g_1~
\theta'~]~=~0\quad,
\label{conchi}
\ee
leads to the following solution
\be
\chi~=~\Xi~\tilde\chi~-~g_0~\frac{\tau_E}{T_E}~\theta'~+~g_1~
\theta'\quad .
\ee
In the last line of \bref{canHam}, $\tilde{H}$ and $\tilde{T}$ are the first class combinations of the bosonic constraints:
\bea
\tilde{H}&=&h-\frac{1}{T_E}F\tau_E\theta'
=\frac{1}{2T_E}({p}^2+(T_E)^2X'^2)
-\frac{1}{T_E}\zeta\tau_E\theta'
\label{tildH}\\
\tilde{T}&=&t+F\theta'
=p\cdot X'+\zeta\theta' 
\eea

Now all constraints
 are conserved for any $g_0,\  g_1 $ and $\tilde\chi$. 
From the canonical Hamiltonian \bref{canHam}
we can also read 
that the first class combinations of the fermionic constraints are
\bea
\tilde{F}\equiv  F\Xi \quad.
\eea
They satisfy a closed algebra,
\bea
\{ \tilde{F}_{A,\alpha} ,\tilde{F}_{B,\beta}\} &=&-4[L_{AC}(C\Xi)_{CB,\alpha
\beta}+\left(
(\tilde{F}\tau_E)_{A,(\alpha}\bar{\theta}'_{B,\beta )}
+\tilde{F}_{A,(\alpha}(\bar{\theta}'\tau_E)_{B,\beta )}\right)\nonumber\\
&&+\frac{1}{2}(\tilde{F}\tau_E\Gamma\theta'\cdot(C\Gamma)_{AB,\alpha\beta})
+\frac{1}{2}\tilde{F}\Gamma\theta'\cdot(C\Gamma\tau_E)_{AB,\alpha\beta})
\nonumber\\
&&+\frac{1}{8}(\partial_1 E^1)\bar{\theta}\Gamma\Xi_{A,(\alpha}\cdot
\bar{\theta}\Gamma\tau_3\Xi_{B,\beta)}
]
\delta (\sigma -\sigma ')\quad \label{SDSTFF}
\eea
with the symmetric bracket,  
$A_{(\alpha \beta)}=A_{\alpha \beta}+A_{\beta\alpha}$.
The square of constraints
are set to be zero in the Hamiltonian algebra .
In the right hand side of (\ref{SDSTFF})
 we have introduced combinations of the bosonic first class
constraints
\be
L_{AB}\equiv T_E
\tilde{H}\delta_{AB}-\tau_{E,AB}\tilde{T}
=\frac{1}{2}\Xi^2_{AB}-F\tau_E\theta'\delta_{AB}-F\theta'\tau_{E,AB} \quad .
\ee

The above constraints, $\tilde{F}=L=0$, form a reducible
first class constraint set. 
Instead we can choose an irreducible and covariant set of
first class constraints:
\bea
\left\{\begin{array}{ccl}
\tilde{F}_1&=&\tilde{F}(1+\tau_{3})/2=0\\
L_{11}&=&T_E\tilde{H}-E^1\tilde{T}=0\\
L_{12}&=&-T\tilde{T}=0
\end{array}\right.\quad.
\eea
The point of this irreducibility is that 
``chiral" projection $ \frac{1+\tau_{3}}{2}$ does not commute with
the first class projection operator, $\Xi$:
$\tilde{F}$ are linearly dependent,
$(\tilde{F})_{\alpha}C^{\alpha}\approx 0~$$\leftrightarrow$
$ ~C^{\alpha}=(\Xi)^{\alpha}_{\ \beta}\tilde{C}^{\beta}$.
On the other hand $\tilde{F}_1$ are linearly independent,
$(\tilde{F}_1)_{\alpha}C^{\alpha}= 0~ $$\leftrightarrow$$~C^{\alpha}=0$.
$\tilde{F}_1$ is the canonical generator of the local supersymmetry 
(kappa symmetry).
The canonical projection operator $\Xi$ in (\ref{Xi}) is related to 
the normalized projection operator $\gamma^{(1)}$
derived by Schwarz \cite{Shgf} as
\bea
\Xi
=[1-(\frac{\sltp~\slPi_1 }{TG_{11}}\tau_1
-\frac{E^1}{T}i\tau_2)](-T\slPi_1 \tau_1)~=~
[1-\gamma^{(1)}](-T\slPi_1 \tau_1) \label{DSpro}
\quad.
\eea

Fermionic constraints independent of the first class constraints 
$\tilde F_1$ are second class;
\bea
F_2 \equiv  F \frac{1-\tau_{3}}{2}=0\quad.
\eea
The Dirac bracket defined by the second class constraints is
\be
\{ A,B\}_D= \{ A,B\} +\{ A,F_2\}\frac{\Xi_{22} C^{-1}}
{2T^2G_{11}}
\{ F_2,B\}\quad.\label{D25}
\ee
Using the Dirac bracket, (\ref{D25}), the algebra of the first class 
constraints is calculated:
\bea
\{ \tilde{F}_{1,\alpha} ,\tilde{F}_{1,\beta}\}_D 
&=&-4[L_{1A}(C\Xi)_{A1,\alpha\beta}+\frac{1}{2}E^1
(2\tilde{F}_{1,(\alpha}\bar{\theta}'_{1,\beta)}
+\tilde{F}_1\Gamma\theta_1'\cdot(C\Gamma)_{\alpha\beta})\nonumber\\
&&~~~~~~~~~~~~~~~~
+\frac{1}{2}T
(2\tilde{F}_{1,(\alpha}\bar{\theta}'_{2,\beta)}
+\tilde{F}_1\Gamma\theta_2'\cdot(C\Gamma)_{\alpha\beta})
]
\delta (\sigma -\sigma ')\label{FF}\\
\{ \tilde{F}_{1,\alpha} ,L_{11}\}_D 
&=&-E^1(\tilde{F}_1(\sigma)+\tilde{F}_1(\sigma'))\delta'(\sigma -\sigma ')
\label{FL1}\\
\{ \tilde{F}_{1,\alpha} ,L_{12}\}_D &=&-T(\tilde{F}_1(\sigma)+\tilde{F}_1(\sigma'))\delta'(\sigma -\sigma ')
\label{FL2}\\
\{ L_{11} ,L_{11}\}_D 
&=&[-2E^1(L_{11}(\sigma)+L_{11}(\sigma'))
+T(L_{12}(\sigma)+L_{12}(\sigma'))]\delta'(\sigma -\sigma ') \label{L1L1}\\
 \{ L_{11} ,L_{12}\}_D 
&=&-T(L_{11}(\sigma)+L_{11}(\sigma'))\delta'(\sigma-\sigma') \label{L1L2}\\
\{ L_{12} ,L_{12}\}_D 
&=&-T(L_{12}(\sigma)+L_{12}(\sigma'))\delta'(\sigma-\sigma')\label{L2L2}
\eea
Here in right hand sides 
the square of constraints 
as well as
the Gauss law constraint are omitted for simplicity.

%%%%%%%%%%%%%%
So far we have assumed that $G_{11}$ is non-vanishing.
If there were a solution with $G_{11}=0$,
 the above discussions of covariant and
irreducible separation of the first and second class constraints
could not be applied.
The Dirac bracket \bref{D25} becomes ill-defined
because 
$G_{11}$ enters in the denominator of the second term.
Furthermore $G_{11}=0$ is inconsistent with the static gauge,
because it represents 
point-like state. 
This situation, $G_{11}=0$, occurs in the ground state of the Green-Schwarz superstring
which is an obstacle
of the covariant quantization.
The condition, $G_{11}\ne 0$, may be described as follows:
Assuming that the inside of the square root of the DBI action
(\ref{sdact}) is non negative 
\be
-G-{\cal F}_{01}^2 \geq 0\quad ,
\ee
then in the conformal gauge, $G_{00}+G_{11}=G_{01}=0$, 
\be
G_{11}^2\geq {\cal F}_{01}^2 \geq 0\quad  .\label{217}
\ee 
Therefore non-vanishing ${\cal F}_{01}$ could avoid $G_{11}=0$ singularity. 
Since ${\cal F}_{01}$ is related to the $E^1$ which is the canonical 
conjugate of $A_1$ by
\be
E^1\equiv T\frac{{\cal F}_{01}}{\sqrt{-G_F}}\quad ,\label{E1F}
\ee
vanishing ${\cal F}_{01}$ means that the electric field, $E^1$, is zero.
Actually $E^1=0$ states are identified with the states 
of the Green-Schwarz superstring 
as will be discussed in section 5. 
In other words non-vanishing value 
of $E^1$ distinguishes the D-string from the Green-Schwarz superstring.

\vskip 6mm
%%%%%%%%%%%%%%%%%%%%%%%%%%%%%%%%%%%%%%%%%%%%%%%%%%%%%%%%%%%%%%%%%%%
\section{Covariant quantization of the super-D-string}
\indent

We take the irreducible and covariant set of constraints 
and construct the BRST charge of the D-string. 
Using with the first class algebra (\ref{FF})
-(\ref{L2L2}) 
we obtain
\be
Q_B= \ \ Q_{B,0}+Q_{B,1}+Q_{B, {\rm nm}}\label{BRS}
\ee
\bea
Q_{B,0}&=&\int L_{11}c_{11}+L_{12}c_{12}+\tilde{F}_1\gamma_1 +(\partial_1E^1)\mu
\label{BRS0}\\
Q_{B,1}&=&  \int [
b_{11}\left( 2E^1c_{11}c_{11}'+T(c_{11}c_{12}'-c_{11}'c_{12})+2
\bar{\gamma}_1\Xi_{11}\gamma_1 \right) \nonumber\\
&&
+b_{12}\left( T(-c_{11}c_{11}'+c_{12}c_{12}')+
2\bar{\gamma}_1\Xi_{21}\gamma_1 \right) \nonumber\\
&&
+E^1\left(
8(\beta_1\gamma_1)(\bar{\theta}_1'\gamma_1)+2(\beta_1\Gamma\theta_1')\cdot
(\bar{\gamma}_1\Gamma\gamma_1)+(\beta_1\gamma_1)c_{11}'-
(\beta_1\gamma_1')c_{11}\right)\nonumber\\
&&
+T\left(
8(\beta_1\gamma_1)(\bar{\theta}_2'\gamma_1)+2(\beta_1\Gamma\theta_2')\cdot
(\bar{\gamma}_1\Gamma\gamma_1)+(\beta_1\gamma_1)c_{12}'-
(\beta_1\gamma_1')c_{12}\right) ]
 \label{BRS1}\\
Q_{B, {\rm nm}}&=&\int \pi_{g,1A}c_{1A}+\pi_{\psi} \tilde{\gamma}_1
\label{BRS2}
\quad .
\eea
The BRST charge (\ref{BRS})
is shown to be nilpotent; $
\{ Q_B,Q_B\}_D~=~0.
$
Since we take the irreducible set of constraints 
infinite number of additional fields are not required in the covariant
formalism.
Analogous to the super-D-particle case (\ref{gfp}), 
the path integral is given by,
\bea 
Z&=&\int d\mu_{\rm can}~\delta(F_2)(\det\{F_2,F_2\})^{-1}\ 
\nonumber\\
&&\exp i\left[\int  
\left(p\dot{X} +\zeta_A\dot{\theta}_A+E^1\dot{A}_1
+b_{1A}\dot{c}_{1A}+\beta_1\dot{\gamma}_1
- \{ \Psi ,Q_B\}_D \right)\right] \quad,
\nonumber\\ 
&&~d\mu_{\rm can}={\cal D}[\zeta_A , \theta_A,  p, X,A_1,E^1;
c_{1A},b_{1A},\gamma_1,\beta_1]\ 
 \quad .
\eea

The covariant gauge condition $\theta_1=0$ \cite{Shgf} can be chosen 
by using local supersymmetry degree of freedom.
In order to make reparametrization gauge degree of freedom manifest
we keep values of $g_\mu$ unfixed and construct 
the gauge fixing function as
\be
\Psi=\int  \frac{g_0}{2T_E}b_{11}+(\frac{g_0}{2}\frac{E^1}{T_E}-
\frac{g_1}{2})\frac{1}{T}b_{12}
  -\beta_1 \psi-\tilde{\beta}_1\theta_1 -\nu A_0 \quad .\label{gfpsi}
\ee
The contribution from the second class constraints 
is also exponentiated by introducing  $\chi_2$ and $\eta_2$. 
The path integral becomes
\bea
Z_{\rm }=\int  d\mu_{\rm can}{\cal D}[ \chi_2, \eta_2, \rho_2; g,\psi,
\pi_{\psi},A_0,\tilde{\gamma},\tilde{\beta}]\ \exp i\int {\cal L} 
\eea
\bea 
{\cal L}&=&F_2\chi_2 +\bar{\eta}_2\frac{(\Xi_{22})C}{2T^2G_{11}}\rho_2
+p\Pi_0+\zeta_1\dot{\theta}_1+F_2\dot{\theta}_2+E^1\dot{A}_1\nonumber \\&&
-g_0\tilde{H}-g_1\tilde{T}
+\tilde{F}_1\psi+(E^1)'A_0
-(F_2-\zeta_2-\bar{\theta}_2\slp)\dot{\theta}_2+\pi_{\psi}\theta_1
\nonumber\\&&
+b_{1A}\dot{c}_{1A}-g_0 h_{gh}
-g_1 t_{gh}
+\beta_1\dot{\gamma}_1+\beta_1\tilde{\gamma}_1+
\tilde{\beta}_1\Xi_{11}\gamma_1 \quad,
\label{ZDst0}
\eea
with
\bea
h_{gh}&=&\frac{E^1}{2T_E}\{2(b_{11}'c_{11}+2b_{11}c_{11}')+b_{12}'c_{12}+
2b_{12}c_{12}'\}
\nonumber \\
&&+\frac{T}{2T_E}(b_{11}'c_{12}+2b_{11}c_{12}'+b_{12}'c_{11}+2b_{12}c_{11}') 
\nonumber \quad,\\
t_{gh}&=&-\frac{1}{2}
(b_{11}'c_{11}+2b_{11}c_{11}'+b_{12}'c_{12}+2b_{12}c_{12}')
\quad .\label{ZDst1}
\eea
Integrating out auxiliary fermionic fields and canonical momenta 
it is expressed with the minimal fields measure 
\be
d\mu={\cal D} [X,A_0,A_1,\theta_2,c_{1A}, {b}_{1A}]\quad.
\ee
The resultant path integrals are
\bea
Z&=&\int d\mu\ {\cal D} [p,E^1;g_0,g_1]\nonumber\\
&&~~~~~~\ \exp \left[i\int \left(\hat{\tilde{p}}\hat{\Pi}_0+E^1\hat{{\cal F}}_{01}-g_0\hat{h}-g_1\hat{t} +{b}_{1A}\dot{c}_{1A}
-g_0h_{gh}-g_1t_{gh}\right)\right] \label{ZDst2}\\
&=&\int d\mu\ {\cal D} [E^1;g_0,g_1]\nonumber\\
&&~~~~~\ \exp \left[i\int \left(-\frac{T_E}{2g_0}(\hat{\Pi}_0-\frac{g_1}{2}\hat{\Pi}_1)^2
-\frac{g_0}{2}T_E\hat{G}_{11} +E^1\hat{{\cal F}}_{01}+{b}_{1A}\dot{c}_{1A}-g_0h_{gh}-g_1t_{gh}
\right)\right] \nonumber\\&&
\label{ZDst3}\\
&=&\int  d\mu\ {\cal D}[ E^1]
\ \exp \left[i\int \left(-T_E\sqrt{-\det \hat{G}_{\mu\nu}}+E^1\hat{{\cal F}}_{01}
 +{b}_{1A}\dot{c}_{1A}\right)\right] \label{ZDst4} \\ 
&=&\int  d\mu
\ \exp \left[i\int \left(-T\sqrt{-\det (\hat{G}_{\mu\nu}+\hat{{\cal F}}_{\mu\nu})}
 +{b}_{1A}\dot{c}_{1A}\right)\right] \label{ZDst5} 
 \quad .
\eea
Where hat variables stand for $\theta_1=0$, $\hat{\cal O}={\cal O}|_{\theta_1=0}$;
 for example,
$
\hat{\Pi}_{\mu}=\partial_{\mu}X-\bar{\theta}_2\Gamma\partial_{\mu}\theta_2 \quad$ etc. 
In the static gauge, reparametrization ghosts disappear and the gauge fixed action is obtained as expected by Schwarz \cite{Shgf}.

It is also interesting to perform $A_{\mu}$ integral 
in the action (\ref{ZDst4}).
The variation of U(1) gauge fields leads to that $E^1$ is constant
\bea
I=\int \left(-T_E\sqrt{-\det {\hat{G}}_{\mu\nu}}-E^1\hat{\Omega}_{01}(\tau_3) +{b}_{1A}\dot{c}_{1A}\right) \quad.\label{IAIA}
\eea
This form is similar to the GS action in the classical level.
In eq.\bref{IAIA} coefficients of the DBI action and 
the Wess-Zumino term, $\Omega(\tau_3)$,
 are
different contrasting with the GS action in which both coefficients
are $T$.
Same coefficients follow from the local supersymmetry invariance,
while different coefficients represent that the local supersymmetry
is fixed.
Corresponding to that the tension of the GS superstring is $T$, 
the tension of the super-D-string is $T_E$.
The coefficient of $\Omega(\tau_3)$ for the super-D-string is $E^1$
and the one for the GS superstring is $1$ in unit of $T$,
so this form may manifest the NS charge coupling.

The gauge fixed action in the conformal gauge is obtained by choosing
$g_0=1$ and $g_1=0$ in the gauge fixing function \bref{gfpsi},
\bea
I&=&\int p\cdot(\dot{X}-\bar{\theta}_2\Gamma\dot{\theta}_2)
-\frac{1}{2T_E}(p^2+T_E^2X'^2)\nonumber\\
&&+E^1(F_{01}-\bar{\theta}_2\slX'\dot{\theta}_2
+\bar{\theta}_2\Gamma\dot{\theta}_2\bar{\theta}_2\Gamma{\theta}_2')
+{\cal L}_{gh}\quad .\label{gaugec}
\eea
Once the gauge fixed action is obtained,
 equations of motion and boundary conditions are determined. 
Although it was supposed to choose the conformal gauge,
obtained field equations are not conformal.
The reasons are the global SUSY, existence of $E^1$
and the covariant gauge, $\theta_1=0$.
One can check explicitly
that the contribution of $\theta_1=0$ breaks conformal form 
in the canonical equation of motion.

Boundary conditions on the bosonic 
coordinates are determined by 
the surface term of the variation of the gauge fixed action \bref{gaugec}:
In the tangential directions of the brane they are
\bea
~~~~~\delta X_{\mu}|_{\rm at\ boundary}=0
\eea
in order to be consistent with the static gauge.
In the transverse direction it may be chosen as
\bea
~~~~~X_i'|_{\rm at\ boundary}=0\quad 
\eea
from the momentum conservation.
On the other hand fermionic coordinate, $\theta_2$,
must satisfy the periodic (antiperiodic)
 condition in the $\theta_1=0$ gauge.

%%%%% Susy Charge %%%%%%%%%%%%%%%
\medskip
\medskip
This system has N=2 global supersymmetry whose charges
are obtained as
\bea
Q_A{\epsilon}_A=\int d\sigma [\zeta\epsilon
-\bar{\theta}\slp\epsilon+\bar{\theta}\slX'\tau_E\epsilon
-\frac{1}{6}\bar{\theta}\Gamma\tau_E\theta'\cdot\bar{\theta}\Gamma\epsilon
-\frac{1}{6}
\bar{\theta}\Gamma\theta'\cdot\bar{\theta}\Gamma\tau_E\epsilon]\quad .
\eea
The global supersymmetry charges make the following algebra:
\bea
\{Q_{A,\alpha},Q_{B,\beta}\}_D
=-2(C\Xi_G)_{A,\alpha\ B,\beta}
+\int (\partial_1 E^1)
\left(\begin{array}{cc}
0&
\bar{\theta}_1\Gamma_{\alpha}\cdot\bar{\theta}_2\Gamma_{\beta }\\
-\bar{\theta}_2\Gamma_{\alpha}\cdot\bar{\theta}_1\Gamma_{\beta }&
0
\end{array}\right)\label{gSUSY}
\eea
where
\bea
(\Xi_G)_{AB} ~\equiv ~\slP-\int \slX'(\tau_E)_{AB}
&=&\slP
\left(\begin{array}{cc}
1&0\\0&1
\end{array}\right)
-\int \slX'
\left(\begin{array}{cc}
E^1&T\\T&-E^1
\end{array}\right) 
\label{Xig}
\eea
and $P\equiv \int p$ is the total momentum.
The second term in (\ref{gSUSY})
is the Gauss low constraint, (\ref{Gauss}). 
We assumed that the surface term vanishes by the boundary condition.
It is important to notice that $E^1$ is constant by the equation 
 of motion and the constraint (\ref{Gauss}) so $\tau_E$ 
is also constant. 
According to Polchinski \cite{Pol}, the SUSY algebra shows
the NS-NS charge and the R-R charge of the super-D-string as 
$(E^1,1)$. 
This is consistent with the result 
obtained by SL(2,Z) representation of the tension
\cite{Shgf}.

There is a bases in which the supersymmetry algebra (\ref{gSUSY})
 is diagonalized
\bea
\left\{\begin{array}{ccl}
\tilde{Q}_1&=&
\frac{1}{\sqrt{2T_E(T_E-E^1)}}(TQ_1+(T_E-E^1)Q_2)\Gamma_1C^{-1}\\
\tilde{Q}_2&=&
\frac{1}{\sqrt{2T_E(T_E-E^1)}}(TQ_2-(T_E-E^1)Q_1)\Gamma_1C^{-1} 
\end{array}\right.\quad,
\eea
in a rest frame where $\int C \slX'=l C\Gamma_1$. Anti-hermiticity of 
$C\Gamma_1$ leads to imaginary  eigenvalues.
The SUSY algebra becomes 
\bea
\{\tilde{Q}_A,\tilde{Q}_B\}_D
=-2i
\left(\begin{array}{cc}
P_0-lT_E&0\\
0&P_0+lT_E
\end{array}\right)\quad.
\eea
Analogous to the previous section, 
the canonical quantization procedure replaces the Dirac bracket 
by the quantum operator bracket;
$\{ q,p\}_D=1\rightarrow [q,p]=i$,
and the norm of the quantum states leads to the BPS bound:
\be
\parallel \tilde{Q}_1|\ast \rangle \parallel^2=\frac{1}{2}\langle \ast|[\tilde{Q}_1,\tilde{Q}_1]_+|\ast\rangle
=(P_0-lT_E)\parallel |\ast \rangle \parallel^2\geq  0 \quad .
\ee
For the BPS saturated state the SUSY algebra would be
\bea
[\tilde{Q}_{A},\tilde{Q}_{B}]
=
\left(\begin{array}{cc}
0&0\\
0&4T_E l
\end{array}\right)
 \quad .
\eea
This shows that one half of the global SUSY is spontaneously broken.
 
For the super-D-string states, the mass operator
comes from the zero-mode of constraints $\tilde H$ in eq.(\ref{tildH}), 
\bea
\int d\sigma \tilde{H}(\sigma)=
\frac{1}{2T_E}[\frac{l}{l^2}P^2 +l(T_E)^2+{\rm higher\ mode}]=0\quad.
\eea 
The higher mode contribution is positive semi-definite 
which would be obtained by using remaining
 constraints. This condition leads to 
\bea
\frac{l}{l^2}P^2 +l(T_E)^2\leq 0 \quad \rightarrow\quad
P_0\geq T_El\quad, 
\eea
i.e. the ground state of the super-D-string is BPS saturated state.
Then the ground state of super-D-string is a 
representation of the half of the N=2 global supersymmetry.

The superficial breaking down of one half of the global SUSY is also caused by the gauge fixing condition, and which is recovered by the non-linear SUSY transformation analogous to the super-D-particle case (\ref{hfSUSY}).
 The new Dirac bracket is introduced
by $F_1=F_2=\theta_1=0$,
\bea
\{ A,B\}_{D2}&=& \{ A,B\}_D 
-\{ A,\theta_1\}\{ F_1,B\}-\{ A,F_1\}\{ \theta_1,B\}\nonumber\\
&&+\{ A,\theta_1\}(\Xi_{12})(\Xi_{22})^{-1}\{ F_2,B\}
+\{ A,F_2\}(\Xi_{22})^{-1}(\Xi_{21})\{ \theta_1,B\}
\eea
where $\{,\}_D$ is defined in (\ref{D25}). 
The non-linear transformation rules generated by $Q_1$ are given by
\bea
\left\{\begin{array}{ccccl}
\delta\theta_2&=&\{\theta_2,Q_1\epsilon_1\}_{D2}&=&
\Xi_{22}^{-1}\Xi_{21}\epsilon_1\\
\delta X&=&\{X,Q_1\epsilon_1\}_{D2}&=&
\bar{\theta}_2\Gamma\Xi_{22}^{-1}\Xi_{21}\epsilon_1
\end{array}\right.\quad .
\eea
The Dirac * variables are calculated as 
\be
Q_1^*=Q_1-\int\tilde{F}_1\frac{\tilde{\slp}-E^1\slPi_1}{-T^2G_{11}}
 ~~~~~,~~~~Q_2^*=Q_2\quad.
\ee
Under the covariant gauge, $\theta_1=0$, $Q_2$ is realized usually while $Q_1$ is realized by using with the local supersymmetry, $\tilde{F}_1$. 
 The global SUSY parameters are related as
\be
\epsilon_2=\Xi_{22}^{-1}\Xi_{21}\epsilon_1=
\left(\frac{E^1}{T}+\frac{\sltp\slPi_1}{TG_{11}}\right)\epsilon_1\quad,
\ee
if they are constants.
Then the true degree of freedom of the global SUSY becomes one half.
In fact the ground state of the D-string with $G_{11}\ne 0$,
N=2 reduces to N=1.
This is consistent with the BPS condition.

%%%%%%%%%%%%%%%%%%%%%%%%%%%%%%%%%%%%%%%%%%%%%%%%%%%%%%%%%%%%%%%%%%%%%%%%%
\section{Relation between the super-D-string and 
the Green-Schwarz superstring}
\indent

The constraint set of the Green-Schwarz superstring 
has the same form as those of the D-string,
(\ref{LH}), (\ref{LT}) and (\ref{F}).
The former is obtained by replacements
$\tau_E\rightarrow T_E\tau_3$ and  $T_E\rightarrow T$. 
At first let us analyze the procedure of $\tau_E\rightarrow T_E\tau_3$ systematically. 
We consider a U(1) rotation, $U=e^{i\tau_2\frac{\psi}{2}}$
, mixing $\tau_1$ and $\tau_3$: 
\bea
\begin{array}{ccl}
\tau_3&\rightarrow&  U^t\tau_3U=\tau_3\cos\psi +\tau_1\sin\psi\equiv \tau(\psi)
\\
\tau_1&\rightarrow&U^t\tau_1U=\tau_1\cos\psi-\tau_3\sin\psi=\tau(\psi+
\frac{\pi}{2})
\end{array}\quad .\label{51}
\eea
We define the $\psi$-dependent
 Wess-Zumino term, $\Omega_{(2)} (\psi)$
\bea
\Omega_{(2)} (\psi)\equiv \bar{\theta}\tau(\psi)\Gamma_md\theta(dX^m+\frac{1}{2}\bar{\theta}\Gamma^md\theta) \quad,\label{52}
\eea
which is a surface term of a super symmetric closed three form,
$I_{(3)}(\psi)~=~d~\Omega_{(2)} (\psi)$ for any $\psi$.
The Wess-Zumino action for the GS superstring is 
$\Omega_{(2)}(0)$, 
while the one for the super-D-string is $\Omega_{(2)}(\pi/2)$.
We can construct one parameter family of models with the
same symmetry as the super-D-string,
whose action is 
\bea
 I&=&-T\int d^2\sigma \sqrt{-{\rm det}(G+{\cal F})}+
T\int \Omega_{(2)}(\psi+\pi/2)\nonumber\\
\label{actpsi}
\\
{\cal F}&=&dA+
%\frac{\partial}{\partial\psi}
\Omega_{(2)} (\psi)\quad .\nonumber
\eea
It is invariant under the local supersymmetry (kappa symmetry)
for any value of $\psi$, in addition  to the reparametrization invariance 
and the super Poincare symmetry. 
The super-D-string is the case for  $\psi=0$ and 
${\cal F}\ne0$ and  
the GS superstring is the case for $\psi=-\pi/2$ and ${\cal F}=0$.

The $U$ transformation is also interpreted as 
the SO(2) rotation of two spinors $\theta_1$ and $\theta_2$,
$\theta_A\rightarrow U_{AB}\theta_B$.
The boundary condition $\theta_1=0$ for the super-D-string is
obtained by $\psi=\pi/2$ rotation from the boundary condition of the
GS superstring, $\theta_1-\theta_2=0$.
For the super-D-string right(left) handedness of fermionic coordinates 
are mixed because of the central extension of the SUSY algebra.
The SO(2) rotation with $\psi_0=\tan^{-1} (T_E-E^1)/T$ makes
fermionic coordinates $\theta_A$ to be right/left handed eigenstates;
in the constraint algebra 
(\ref{SDSTFF}) the bosonic constraint $L_{AB}$
becomes diagonal.
%\bea
%\{\tilde{F}_A,\tilde{F}_B\}\approx -4L_{AB} C\Xi
%\rightarrow -4(ULU^t)_{AC}C(U\Xi U^t)_{CB}  \quad,
%\eea 
The SO(2) rotates the global SUSY algebra also,
%\bea
%[Q_A,Q_B]\approx -2C(\Xi_G)_{AB}
%\rightarrow
%-2C(U\Xi_GU^t)_{AB}\quad ,\label{QQQQ}
%\eea
so it seems that SO(2) relates
the NS-NS charge and the R-R charge 
when they are identified with the diagonal element and off-diagonal element of SUSY central charges respectively. This degeneracy occurs due to the flatness
of the target space. 
By taking into acccount background with a dilaton $\phi$ and an axion $\chi$, the SUSY central charges becomes SL(2,R) covariant representation. For the flat background the SUSY central charges for a unit string length is essentially
\bea
(\tau_E)_{AB}~=~(E^1~,~T) \pmatrix{\tau_3 \cr \tau_1}_{AB}
\eea
in \bref{Xig}. With the background it becomes 
\bea
(\tau_{E,{\rm bg}})_{AB}&=&(E^1~,~T) e^{\phi/2}
 \pmatrix{1&0\cr-\chi&e^{-\phi}}  \pmatrix{\tau_3 \cr \tau_1}_{AB} \nonumber \\
&\equiv &(q_{NS}~,~q_{R})~K~\pmatrix{\tau_3 \cr \tau_1}_{AB},\label{qKt}
\label{taubg}
\eea
where $K$ is a representation of SL(2,R)/SO(2) satisfying
\bea
&&K^{T}\CM K=1 \quad,\quad \CM=e^{\phi}\pmatrix{|\lambda|^2&\chi\cr \chi&1}
\quad. 
%\nonumber\\
%&&\CM\rightarrow \Lambda\CM\Lambda^T\quad,\quad K\rightarrow\Lambda K
%O\quad
%for \quad \Lambda\in SL(2,R),~~O\in SO(2) 
\eea
Therefore $K\pmatrix{\tau_3 \cr \tau_1}$ transforms as a SL(2,R) doublet 
when $\pmatrix{\tau_3 \cr \tau_1}$ does as a SO(2) doublet. 
$K\pmatrix{\Omega(\psi) \cr \Omega(\psi+\pi/2)}$ corresponds to 
$\pmatrix{B_{\mu\nu}^{NS} \cr B_{\mu\nu}^{R}} $ in the Green-Schwarz formalism. The $\tau_{E,{\rm bg}}$ in \bref{taubg} is SL(2,R) covariant if we regard
$(E^1~,~T)$ as a SL(2,R) (contravariant) doublet.

The relation between the action \bref{actpsi} and the manifest S-dual 
action \cite{Twsl} is interesting. 
It is introduced a "dual U(1) field $\tilde A$ " as a SL(2,R) partner
of the U(1) field $A$, and  $\tilde{E}^1$ plays a role of $~T$ 
in \bref{taubg}. 
The action \bref{actpsi} is flat case of
manifest  the S-dual action \cite{Twsl}.
Indeed the combination of the momenta $\tilde{p}$ in \bref{3.7} is a remnant of SO(10) $\times$ SL(2,R) as (2.8) in \cite{Twsl}.

\medskip
Next let us examine the condition $T_E\rightarrow T$ for the super-D-string 
to coincide with the GS superstring. 
The difference between the Green-Schwarz superstrings and the super-D-string 
is the value of $E^1$ in the constraints set \bref{LH}, \bref{LT} and \bref{F} 
, while it is the value of ${\cal F}_{01}$ in the action level \bref{actpsi}.
Since $E^1$ is the canonical conjugate of $A_1$ as \bref{E1F},
both $E^1=0$ and ${\cal F}_{01} =0$ are same conditions.
 As seen in (\ref{217})
these conditions allow existence of
the point-like ground state. 
There is another situation for the super-D-string 
where the above condition
is approximately satisfied;
There are so highly excited states of 
the super-D-string where ${\cal F}$ can be neglected.
These states behave similar to highly excited states of the GS superstring. 
Then the covariantly quantizable super-D-string inspires the 
covariantly quantizable GS superstring.
However at the particle-like state 
each elements of the projection operator in (\ref{Xi})
become nilpotent, $\Xi_{AB}\rightarrow \slp\delta_{AB}$.
Then neither covariant separation of fermionic constraints nor 
well-defined Dirac bracket 
can be constructed.     
The SUSY algebra for the massless point ground state becomes
\bea
\{{Q}_{A},{Q}_{B} \}_D
=-2C\slP \delta_{AB}
\rightarrow
-2P_+
\left(\begin{array}{cc}
C\Gamma_-
&0\\
\noalign{\vskip0.1cm}
0&
C\Gamma_-
\end{array}\right)
 \quad .
\eea
Each element of the above matrix is nilpotent 
and it is difficult to find the covariant Nambu-Goldstone fermions.
The
light-cone variables as Nambu-Goldstone fermions are relatively
convenient.
Therefore the covariant quantization of the Green-Schwarz superstring is still 
open problem because of the massless particle ground state.

\section{Conclusions}
\indent

We performed the canonical analysis of super-D-brane actions for p=0 and 1.
Irreducible and covariant separation of fermionic constraints into the 
first and second classes can be possible thanks to off-diagonal elements 
of the fermionic constraints algebra, $\Sigma$ in (\ref{Sigma}) for p=0 and 
$\Xi$ in (\ref{Xi}) for p=1. 
Off-diagonal elements give mass to 
ground states of D-branes.
In the canonical analysis of the super-D-string 
the singularity at $G_{11}=0$ appears
where the covariant separation and the Dirac bracket become ill-defined.
A sufficient condition 
to avoid this singularity is ${\cal F}\ne 0$.
This condition also guarantees 
static picture of it. 
Non-zero ${\cal F}$ means
the U(1) excitation on the world sheet of the D-brane (string) which is
caused by strings whose ends are on the brane. 
The physical situation corresponding to ${\cal F}=d{\cal A}+\Omega_{(2)}\ne 0$ 
should be clarified more. 

The BRST charges for the super-D-particle and the super 
D-string are calculated 
and covariant gauge fixed actions are derived. 
Since we have an irreducible set of constraints
infinite number of fields are not required in contrast to the
covariant quantization of the superparticle and the superstring theories.
Equations of motion are examined after gauge fixing:
For the super-D-particle $X$ and $\theta$ satisfy free field 
equations, but
 the gauge fixed action can not be written in free form because 
of the manifest global supersymmetry. 
For the super-D-string  
$X$ and $\theta$ do not satisfy conformal equations
not only because of the central extension of the SUSY algebra and 
existence of $E^1$ 
but also because of the covariant gauge $\theta_1=0$.

The global supersymmetry algebra leads to the BPS bound.
The super-D-particle and the ground state of the super-D-string are 
BPS saturated states.
In the BPS saturated state, 
one half of the global supersymmetry is spontaneously broken
while another half is 
trivial.
The point is that this BPS condition can be written covariantly,
so that the Nambu-Goldstone fermion is covariant spinor. 
This covariance reflects covariance of the gauge fixing.
In the action one half of the global SUSY is realized manifestly,
while another half is realized trivially.  
The use of the Dirac stared variable clarified 
that trivial global SUSY generator is nothing but 
the local SUSY (kappa symmetry) generator.

As results of checking physical picture of the super-D-string action 
(\ref{sdact}) for ${\cal F}\ne 0$ ($E^1\ne 0$) :
(1) It has the NS-NS charge and the R-R charge, 
$(q_{NS},q_{R})=(E^1,1)$. 
(2) The ground state of the super-D-string is the BPS saturated state. 
(3) The D-string tension is scaled as 
$T\rightarrow T_E=\sqrt{(E^1)^2+T^2}=\sqrt{q_{NS}^2+T^2q_{R}^2}$. 
(4) It is consistent with the static gauge.

We also clarified the relation between the Green-Schwarz superstring and the super 
D-string.
It turns out that the local supersymmetric D-string action can be parametrized 
by a SO(2) parameter, $\psi$, as (\ref{actpsi}).
The GS superstring is the case for $\psi=-\pi/2$ with ${\cal F}=0$, 
while the super-D-string is the case for $\psi=0$ with ${\cal F}\ne 0$. 
This SO(2) mixes up the right/left handedness,
 so it changes the boundary condition of spinor coordinates. 
This SO(2) also acts on the global SUSY algebra. 
By taking into account a dilaton and an axion, the SUSY central charges become SL(2,R) representation. 
In contrast with an approach to realize SL(2,R) manifestly \cite{Twsl},
the action 
\bref{sdact} does not have manifest SL(2,R) symmetry since the dual U(1) field is fixed. 
The action \bref{sdact} describes superstrings with charges 
$(q_{NS},q_{R})=(E^1,1)$ except $E^1=0$, where singularity appears then the string reduces to a fundamental superstring.
Therefore we suggest that rather the value of 
$E^1$ plays a role of the order parameter of the action \bref{sdact};
the super-D-string exists for $E^1\ne 0$ 
while the fundamental GS superstring exists for $E^1=0$. 
Further study is necessary for the super-D-string physics.

\medskip\noindent
{\bf Acknowledgements}\par
\medskip\par
The authors would like to thank 
Joaquim Gomis for helpful discussions.
M.H. is partially supported by the Sasakawa Scientific Research Grant from the Japan Science Society.
%\appendix

%\newpage

\end{document}